\documentclass[doublecol]{epl2} 
\usepackage{graphicx} \usepackage{amsmath} \usepackage{amssymb}
\newcommand{\BEQ}{\begin{equation}}
\newcommand{\EEQ}{\end{equation}}
\newcommand{\BEA}{\begin{eqnarray}}
\newcommand{\EEA}{\end{eqnarray}}

\renewcommand{\d}{{\rm d}}

\newcommand{\fm}{\widetilde{F}}
\newcommand{\sd}{{{\bf d}}}

\title{Thermodynamics of adiabatic feedback control.}

\author{A.E. Allahverdyan and D.B. Saakian}
\institute{Yerevan Physics Institute,
Alikhanian Brothers Street 2, Yerevan 375036, Armenia.}


\abstract{ We study adaptive control of classical ergodic
Hamiltonian systems, where the controlling parameter varies slowly in
time and is influenced by system's state (feedback).  An effective
adiabatic description is obtained for slow variables of the system. A
general limit on the feedback induced negative entropy production is
uncovered. It relates the quickest negentropy production to fluctuations
of the control Hamiltonian. The method deals efficiently with the
entropy-information trade off.  }

\begin{document}

\maketitle

{\it Introduction.} Relations between the control theory and physics
have a long history.  The notorious Maxwell's demon, conceived yet in
XIX'th century, is in fact a control device that aims to reduce the
entropy of a statistical system \cite{demon,pop}. Founders of
cybernetics recognized the entropy reduction as one of the basic goals
of control \cite{cyber,cyber_ash,leon}.  This became even more important once
it was understood that the statistical description and thermodynamic
relations are needed not only for the macroscopic situation, but also
for few-body chaotic and stochastic systems \cite{munster,berdi,sasa,rugh}.
Indeed, the first attempts of relating entropy and information during
control operations were made in the early days of cybernetics
\cite{cyber_ash,leon} and where based on thermodynamics; see
\cite{demon} for a fuller historical perspective. More recent theories
of entropy-information-control relationship were presented in
\cite{pop,tou}. The results of \cite{tou} found applications in the
theory of chaotic systems control, where the entropy reduction is again
the basic goal. This field has undergone an explosive
development due to synthesing the physical and control scientific ideas
\cite{rev,bech,fradkov}. 

Much attention was devoted recently to the control of Brownian systems
(mesoscopic particles coupled to a thermal bath)
\cite{kittel}--\cite{udo}. This field is expected to have wide
applications in various areas of nanoscience.  The first theory of
feedback driven cooling (entropy-reduction) of a Brownian particle was
developed within statistical thermodynamics \cite{kittel}. This theory
plays an important role for recent experimental cooling schemes in
nano-physics \cite{exp,atomic_micro,se_nature}, e.g., in atomic force
microscopy \cite{atomic_micro}. In a related context, Ref.~\cite{cohen}
studied experimentally how the feedback control applied to a Brownian
nanoparticle generates forces of rather general shape, including
non-potential forces. 

Following to the experimental development of feedback cooling methods,
Ref.~\cite{qian} presented the thermodynamic analysis of a Brownian
particle, which couples to a thermal bath and is manipulated by control
fields so that to cool the bath. The authors of \cite{qian} gave a
general recipe for calculating the entropy pumped out of the bath versus
the entropy produced during the operation of the Brownian particle. The
quantum extension of this setup was investigated in \cite{pavon}.
Fluctuation theorems for the classical Brownian control setup were
studied in \cite{qian_PRE}. 

Control of Brownian particles is also employed for generating a directed
motion; see \cite{rev_rat} for reviews.  This task is important for
contructing nanoscale engines (rachets or Brownian motors).  Theoretical
and experimental proposals for feedback driven ratchets appreared
recently in \cite{cao} and \cite{se_nature}, respectively. 

Finally we should mention works devoted to the open-loop (non-feedback)
control of Brownian particles \cite{seki,udo}. These studies are
especially relevant for statistical physics, since the basic
formulations of the second law are in fact control-theoretical
statements \cite{arm}. 

Here we explore an approach to the adaptive feedback control of classical
Hamiltonian systems.  Feedback means that the dynamics of a control
parameter of the Hamiltonian is influenced by system's state, i.e., it
performs an adaptive motion, while in the non-feedback (open-loop)
control the motion of the control parameter is prescribed.  Our main
assumption is that the control parameter moves slowly. Assuming the
ergodicity of certain observables we develop a general thermodynamic
description of the feedback control process. In particular, we find the
control fields that ensure the quickest reduction of the entropy
(noise). This reduction is related to fluctuations of the controlling
part of the Hamiltonian. We also describe the
entropy-information trade-off: how limitations of the information
available on system's state decrease the speed of the entropy reduction
and change the qualitative features of the control process. 

Note that in contrast to the above works on the Brownian particles
control, we shall work with the full Hamiltonian system, and not with a
small particle coupled to a large thermal bath. Moreover, we focus on
Hamiltonian systems with finite degrees of freedom, though the presented
theory applies to macrocopic Hamiltonian systems, e.g., the
particle plus the bath.  The macroscopic
system control will be studied elsewhere \cite{aa}. 

{\it Basics of the method.}
Consider a system with $n$ degrees of freedom and Hamiltonian
$H(p,q,R)$. The equations of motion read
\BEA \label{1}
\dot{q}=\partial_p H(p,q,R), \qquad
\dot{p}=-\partial_q H(p,q,R),
\EEA where $q=(q_1,...,q_n)$ and $p=(p_1,...,p_n)$ are,
respectively, the coordinates and momenta, and where $R$
is a time-varying parameter (the extension to several parameters is
straightforward).
The parameter $R$ is changed externally controlling
the system (the goals of control are indicated below).
The evolution of $R$ is described in the standard way
of the adaptive control \cite{but,bech,fradkov,rev,adaptive}:
\BEA \label{2}
\tau_R\dot{R}=F(E,R,z), \quad z\equiv(q,p), \quad |F|\leq \fm,
\EEA
where $F$ is the control field, $\tau_R$ is a characteristic time of
$R$, and where $E$ is system's energy. The constraint $|F|\leq \fm$ with
constant $\fm$ is a natural condition for practical realizability of the
control setup. Many control tasks get the real physical meaning only
after imposing such constraints on the magnitude of the control fields
\cite{but}. 

As shown by (\ref{2}), the controlling parameter $R$ is subjected to
feedback: the dynamical variables $z$ and $E$ influence,
via suitable engineering, the evolution of $R$.  Control processes without
feedback (open-loop control) correspond to $F=F(R)$, with predetermined
motion of $R$. 

So far we presented a standard and rather general setup for control
problems. In particular, Eq.~(\ref{2}) contains many adaptive control
processes known in literature \cite{but,bech,fradkov,rev,adaptive}. One
of our basic assumptions is that the motion of $R$ is adiabatic, i.e.,
that the time-scale $\tau_R$ of $R$ is much larger than the
characteristic time $\tau_S$ of the system (defined after (\ref{karamba2})). 
Eqs.~(\ref{1}, \ref{2}) show that together with $R$ 
also the energy $E$ becomes a slow variable, 
\BEA
\label{morda}
\frac{\d H}{\d t}=\frac{\d R}{\d t}\, \partial_R H(z,R)=
\frac{1}{\tau_R}F(E,R,z) \partial_R H(z,R),
\EEA
provided that the controlling part of the Hamiltonian $\partial_R H$ is
limited. Note that $\frac{\d R}{\d t}$ is equal to the small parameter
$\frac{1}{\tau_R}$ times the function $F(E,R,z)$, which changes fast
together with $z$.  Thus we need the constraint (\ref{2}) on the
magntitude of $F(E,R,z)$ for the adiabatic assumption to apply.

We want to get from (\ref{1}--\ref{morda}) averaged equations for the
slow variables $E$ and $R$. To this end, define the microcanonic distribution:
\BEA
{\cal M}(z,E,R)=\frac{\delta[E-H(z,R)]}{\partial_E\Omega(E,R)},\\
\Omega(E,R)=\int\d z\, \theta[E-H(z,R)],
\label{quito}
\EEA
and where $\delta(x)$ and $\theta(x)$ are, respectively,
the delta and step function. Here $\Omega(E,R)$ is the phase-space volume
at energy $E$; its derivative $\partial_E\Omega(E,R)$ 
defines the normalization of ${\cal M}(z,E,R)$.

Denote by $z_t$ and $R_t$ the solutions of (\ref{1}, \ref{2}).  On times
$\tau_R\gg\tau\gg\tau_S$ we have from (\ref{1}, \ref{2}) for the slow
derivative ${\sd E}/{\sd \tau}$ :
\BEA
&&\frac{\sd E}{\sd \tau}
\equiv
\frac{1}{\tau}\,[\,H(z_{t+\tau},R_{t+\tau})-H(z_t,R_t)\,]\\
&&=\int_t^{t+\tau}\frac{\d s}{\tau}\,\frac{\d H}{\d s}(z_s,R_s)
=\int_t^{t+\tau}\frac{\d s}{\tau}\,\dot{R}_s\,
\frac{\partial H}{\partial R}(z_s,R_s),
\nonumber\\
&&=\frac{1}{\tau_R}
\int_t^{t+\tau}\frac{\d s}{\tau}\,F(E_s,R_s,z_s)
\frac{\partial H}{\partial R}(z_s,R_s),
\nonumber\\
&&=\frac{1}{\tau_R}
\int_t^{t+\tau}\frac{\d s}{\tau}\,F(E_t,R_t,z_s)
\frac{\partial H}{\partial R}(z_s,R_t)
+o(\frac{\tau}{\tau_R}).
\label{hek}
\EEA
As a consequence of the adiabatic assumption, 
the last integral in (\ref{hek}) refers to 
the dynamics with $R_t=$const and $E_t=$const.
Denote 
\BEA
\label{kuku}
w(z)\equiv F(E_t,R_t,z)\,\partial_RH(z,R_t),
\EEA
and write the time-average in (\ref{hek}) as $\int_t^{t+\tau}\frac{\d s}{\tau}\,w(z_s)$.
Consider the following obvious relation:
\BEA
\label{karamba1}
\int \d z \,w(z){\cal M}(z,E)
=\frac{1}{\tau}\int_t^{t+\tau}\d s
\int \d z \,w(z){\cal M}(z,E).
\EEA
In the RHS of (\ref{karamba1}) we change the integration variable as
$y={\cal T}_{t-s}\, z$, where ${\cal T}_{t}$ is the flow generated by
the Hamiltonian $H(z)$ between times $0$ and $t$.
Employing Liouville's theorem, $\d z=\d y$, and energy conservation, 
${\cal M}(z,E)={\cal M}(y,E)$, one gets
\BEA
(\ref{karamba1})=\int \d y\,{\cal M}(y,E)\,
\frac{1}{\tau}\int_t^{t+\tau}\d s \,w({\cal T}_{s-t}\,y).
\label{karamba2}
\EEA
If $w(z)$ is an {\it ergodic observable} of the $R_t=$const dynamics,
then for $\tau\gg \tau_S$ the time-average $\frac{1}{\tau}\int_t^{t+\tau}\d s \,w({\cal T}_{s-t}\,y)$
in (\ref{karamba2}) 
depends on the initial condition $y$ only via its energy $H(y)$ \cite{munster,berdi,vk}. 
(Thus, $\tau_S$ is the relaxation time of $w(z)$.) In particular, the dependence of the
precise value of $y$ is irrelevant provided that the condition $H(y)=E$ holds. Since
${\cal M}(y,E)$ is a delta-function concentrated at that value of energy,
the integration over $y$ in (\ref{karamba2}) drops out, and we get from 
(\ref{karamba1}, \ref{karamba2}) that the time-average is equal to the
microcanonic average at the energy $E$
\BEA
\label{kaba}
\int \d z \,w(z){\cal M}(z,E)=\frac{1}{\tau}\int_t^{t+\tau}\d s \,w({\cal T}_{s-t}\,y).
\EEA 
Applying this to (\ref{hek})
we get
\BEA
\label{hek1}
&&\tau_R\frac{\sd E}{\sd \tau}
=\int \d z\,\, {\cal M}(z,E,R)\, F(E,R,z)\,
\partial_R H(z,R),~~~~~ \\
&&\tau_R
\frac{\sd R}{\sd \tau}
=\int \d z\,\, {\cal M}(z,E,R)\, F(E,R,z)
\equiv
\langle F\rangle_{E,R},~~~~~
\label{hek2}
\EEA
where (\ref{hek2}) is derived analogously to (\ref{hek1}) by assuming 
the ergodicity of $F(E,R,z)$.

Eqs.~(\ref{hek1}, \ref{hek2}) are the basic equations of the adiabatic
feedback theory. We list again the assumptions employed in their
derivation: {\it i)} $E$ and $R$ are slow variables; {\it ii)}
conservation of energy for $R={\rm const}$; {\it iii)} Liuoville's
theorem; {\it iv)} ergodicity of two phase-space observables: $w(z)$
defined by (\ref{kuku}) and the controlling parameter $F(E,R,z)$. 

Instead of $2n+1$ equations (\ref{1}, \ref{2}) we have in (\ref{hek1}, \ref{hek2})
only two equations for $E$ and $R$. They are autonomous, since they do
not depend on the precise initial value of $z$, provided it was on the
initial energy shell $E_{i}$. Thus the control processes described by
(\ref{hek1}, \ref{hek2}) can operate under conditions, where the initial
values of $z$ are not known or the dependence from them is not desirable.  The
price to be paid for this is that now only functions of $E$ and $R$ can
be controlled.  

Recall that for a (fully) ergodic system all the sufficiently smooth
observables are ergodic, while a non-ergodic system can still have some ergodic
observables; see \cite{vk} for the general theory. Now note that since
no ergodicity of all observables is assumed in deriving
Eqs.~(\ref{hek1}, \ref{hek2}), they apply to some non-ergodic systems.
Another reason for applying (\ref{hek1}, \ref{hek2}) to non-ergodic
systems is that the ergodicity may be restored under small perturbation
\cite{munster,berdi}.  Thus the scheme applies to most of chaotic
systems. 

{\it Definition of entropy.} Entropy and information play important
roles in the control theory: the very possibility of applying feedback
is due to the information available on the state of the system, while
entropy reduction [negentropy production] is necessary for immunizing
the system from sources of noise and instability
\cite{pop,cyber,cyber_ash}. Thus our next task is to obtain from
(\ref{hek1}, \ref{hek2}) the maximal negentropy production rate.  Recall
that for a (partially) ergodic system the entropy is defined as
\cite{munster,berdi}:
\BEA 
S=\ln \Omega(E,R) \equiv \ln\int \d z
\,\theta(E-H(z,R)).
\label{entropy} 
\EEA
Eq.~(\ref{entropy}) satisfies to all desired features of entropy, even
if the number $n$ of the system degree of freedom is finite: 

{\bf 1.} For the temperature $T$ defined via the standard thermodynamic formula
\BEA
\label{dadd}
1/T=\beta=\partial_E\ln \Omega(E)=\frac{1}{\Omega(E)}\int \d z
\,\delta(E-H(z,R)),
\EEA
the integration by parts leads to equipartition \cite{munster,berdi}:
$\langle y\,\partial_y H\rangle=T$, where $y$ is any canonical variable,
while $\langle ...\rangle$ is the average over microcanonic distribution
(\ref{quito}). For the standard Hamiltonian
$H=\sum_{k=1}^n\frac{p^2_k}{2}+V(q_1,..,q_n)$ we get $\langle
p^2_k\rangle=T$ for any $k$, which is the standard form of
equipartition.  Note that $T$ in (\ref{dadd}) is non-negative.

{\bf 2.} $S$ satisfies to the first law of thermodynamics for the
microcanonic ensemble \cite{munster,berdi,campisi_heat}. 

{\bf 3.} $S$ satisfies to two formulations of the second law that
describe a partially ergodic system subjected to an open-loop (i.e.,
non-feedback) control: {\it i)} $S$ remains invariant in an
adiabatically slow [open-loop] process; see
\cite{munster,berdi,sasa,rugh} and the discussion after
(\ref{kabanets}).  {\it ii)} $S$ increases under a non-slow [open-loop]
process, provided that the system starts its evolution from an
equilibrium state \cite{sasa,mwp}. The latter formulation is closely
related to the minimum work principle \cite{arm}. 

In the thermodynamical limit $S$ goes to the more usual
Boltzmann expression \cite{munster,berdi}:
\BEA
\label{boltz}
S_B(E)=\ln [\partial_E\Omega] =\int \d z\, \delta(E-H(z,R)).
\EEA

{\it None} of the above features {\bf 1-3} holds if we apply $S_B$ out
of the thermodynamic limit, e.g., $S_B$ is not constant for open-loop
adiabatic processes. This is because $S$ is the unique adiabatic
invariant for ergodic systems; see \cite{munster,berdi} and the
discussion after (\ref{kabanets}).  Another problem with using the
Boltzmann expression $S_B$ directly for finite systems is that the
temperature defined via (\ref{boltz}) and the standard thermodynamic
formula as $1/T_B=\partial_E S_B(E)$ is in general not even positive
\cite{berdi}. The problems in attempting to use $S_B$ as the proper
entropy will be illustrated below by a concrete example; see the
discussion after (\ref{orojkhon}). 

We are thus convinced that $S$ is the proper expression
of entropy for both finite and macroscopic ergodic systems.

{\it Negentropy production.} We now determine the evolution of the entropy $S$
according to (\ref{hek1}, \ref{hek2}). Using (\ref{hek1}--\ref{dadd}) we get
\BEA \tau_R\frac{\sd S}{\sd \tau} =\beta (E,R)\,\langle\,
F\,\,[\,\partial_R H-
\langle\partial_R H\rangle_{E,R}\,]\,\rangle_{E,R}
\label{ab1},
\label{kabanets}
\EEA
where $\langle ...\rangle_{E,R}$ is defined in (\ref{hek2}), and where
$\beta(E,R)>0$ is the inverse temperature defined in (\ref{dadd}).
Eq.~(\ref{ab1}) shows that if there is no feedback over the fast
variable $z$, $F=F(E,R)$, the entropy $S=\ln\Omega(E,R)$ is conserved on
the slow time, i.e., it is an adiabatic invariant
\cite{munster,berdi,sasa,rugh}. Recall that for ergodic systems
this is the unique adiabatic invariant: any other quantity $K=K(E,R)$
that remains constant for open-loop adiabatic processes is a function of
the phase-space volume $\Omega$: $K=K(\Omega)$ \cite{munster,berdi}. 
Indeed, since in general $\partial_E\Omega(E,R)>0$ we can express (for a fixed $R$)
$E$ as a function of $\Omega$ and $R$: $E=E(\Omega,R)$. Putting this into $K(E,R)$
we re-express it as a function of $\Omega$ and $R$: $K=K(\Omega,R)$. Differentiating $K(\Omega,R)$
over the slow time we get:
\BEA
\label{halt}
\frac{\sd K}{\sd \tau} =\frac{\sd \Omega}{\sd \tau}\,\,\frac{\partial K}{\partial \Omega} +
\frac{\sd R}{\sd \tau}\,\,\frac{\partial K}{\partial R}. 
\EEA
Since both $\Omega$ and $K$ are assumed to be adiabatic invariants, $\frac{\sd \Omega}{\sd \tau}
=\frac{\sd K}{\sd \tau}=0$, we get from (\ref{halt}) that $\frac{\partial K}{\partial R}=0$, i.e.,
$K$ is a function of $\Omega$ only. 

However, for feedback processes the entropy does change. Let us find
the feedback $F(E,R,z)$ that maximizes the negentropy production $-\frac{\sd S}{\sd \tau}$ under
the natural constraint $|F|\leq \fm$. Since the RHS of (\ref{ab1})
is a linear function of $F$, and since $\beta>0$, the extremum is
achieved on the boundaries $|F|= \fm$ of the allowed range.
This implies for the most negative negentropy production $\frac{\sd S}{\sd \tau}$,
which we denote as $\widetilde{\frac{\sd S}{\sd \tau}}$:
\BEA
\label{ab2}
&&\tau_R\widetilde{\frac{\sd S}{\sd \tau}}
=-\beta\, \fm
\langle \,\left|
\partial_R H-
\langle\partial_R
H\rangle_{E,R}
\right|\,
\rangle_{E,R},\\
\label{ooo1} &&F(E,R,z)=-\fm\, {\rm sign}[\,\partial_R H-
\langle\partial_R H\rangle_{E,R}\,].
\EEA
Eq.~(\ref{ab2}) bounds the rate of the entropy reduction, and it is
related to fluctuations of the control Hamiltonian $\partial_R H$ on the
surface of constant energy.  The optimal control function (\ref{ooo1})
is seen to be discontinuous. Note that Ref.~\cite{udo} reports that
discontinuous control fields is a general feature of the optimal
open-loop control that operates in a finite time. We see a
similar effect for a feedback setup, which is not constrained globally
to a finite operation time.

The phase-space volume $\Omega(E,R)$ is a Lyapunov function for the dynamics
described by (\ref{hek1}, \ref{hek2}, \ref{ooo1}), since it is obviously
non-negative, and since it is non-increasing,
$\tau_R\widetilde{\frac{\sd \Omega}{\sd \tau}} \leq 0$, 
as follows from (\ref{ab2}). The non-negativity and 
non-increasing of $\widetilde{\frac{\sd \Omega}{\sd \tau}}$
imply, via the Lasalle principle \cite{lasalle}, that the
long-$\tau$ solutions $\bar{E}, \, \bar{R}$ of (\ref{hek1},
\ref{hek2}, \ref{ooo1}) satisfy $\widetilde{\frac{\sd \Omega}{\sd
\tau}}(\bar{E}, \bar{R})=0$, which leads via (\ref{ab2}) to 
\BEA
\int \d z\, \delta(\bar{E}-H(z,\bar{R}))[\partial_R H(z,\bar{R})- \langle\partial_R
H\rangle_{\bar{E},\bar{R}} ]=0.
\label{kon_tiki}
\EEA
There are two possibilities for
satisfying the equality in (\ref{kon_tiki}): {\it i)} the long-$\tau$
solutions converge to a stable fixed point (i.e., energy minimum) of the
original Hamiltonian system (\ref{1}). This means that the phase-space volume
$\Omega(E,R)$ decays to zero, while the entropy $\ln\Omega(E,R)$ decays
to its minimal value minus infinity. 
{\it ii)} The second possibility of realizing (\ref{kon_tiki}) is that
the long-$\tau$ solutions converge to a point $(\bar{E},\bar{R})$ such
that $\partial_R H(z,\bar{R})$ as a function of $z$ is constant on the
energy shell $H(z,\bar{R})=\bar{E}$, i.e., it is a constant of motion
for the fixed values of $E=\bar{E}$ and $R=\bar{R}$.  Since the second
option is unstable to small perturbation in the control Hamiltonian
$\partial_R H(z,R)$, the
first option is more likely to be realized: the feedback setup
(\ref{ooo1}) drives the system toward an energy minimum, where the
phase-space volume $\Omega$ is equal to zero. 

{\it Limits on the available information.} When obtaining the maximum rate
(\ref{ab2}) of negentropy production we only assumed that the magnitude
$|F|$ of the controlling parameter is limited from above; see (\ref{2}). More crucial
restrictions come into play when noting that in practice the very information
available on the state of the system is limited. 
We thus assume that the full knowledge of $z$ is not available for the controller;
only some function $\Phi(z)$ of $z$ is known, and thus the feedback $F$ in (\ref{2})
depends on $z$ only via $\Phi(z)$: 
\BEA
\label{kentavr}
F(E,R,z)=f(E,R,\Phi(z)). 
\EEA
Note that this implies a reduction of the data $z$, since for simplicity
we take one|in general not one-to-one|function $\Phi(z)$ instead of the
vector $z=(q_1,...,q_n; p_1,...,p_n)$. This is the standard way of
modeling the data reduction in the information theory, known also as
coarse-graining or statistics taking \cite{cover_thomas}.  All the
standard measures of information|e.g., Shannon entropy, relative
entropy, the Fisher information|are known to decrease after taking a
(not one-to-one) function of the data.  In other words, the data
reduction means information decrease with respect to any definition of
information.  In the extreme case, where no information is available for
the feedback we have $\Phi(z)={\rm const}$.  Rewriting (\ref{ab1}) as
\BEA
\tau_R\frac{\sd S}{\sd \tau} =\beta(E,R)\int \d y \,
f(E,R,y)\, \psi(E,R,y), 
\EEA
where we defined 
\begin{gather}
\label{katu}
\psi(E,R,y)\equiv\int\d z\, {\cal M}(z,E,R)\times    \\
\delta[y-\Phi(z)]\,[\,\partial_R H(z,R)-
\langle\partial_R H\rangle_{E,R}\,],\nonumber
\end{gather}
and applying the same derivation as for (\ref{ooo1}, \ref{ab2}), we get
for the most negative $\frac{\sd S}{\sd \tau}$ at the given $\Phi(z)$:
\begin{gather}
\frac{\sd S}{\sd \tau}
=-\frac{\beta \fm}{\tau_R}\int \d y\left|\,\psi(E,R,y)
\,\right|.
\label{ab3}
\end{gather}
This value of $\frac{\sd S}{\sd \tau}$ is achieved for the feedback:
\BEA
\label{karro}
f(E,R,y)=-\fm\, {\rm sign}[\,\psi(E,R,y)\,] ,
\EEA
where $F(E,R,z)$ is recovered from (\ref{kentavr}, \ref{karro}).
Thus the maximal negentropy production under information limitation is
related to fluctuations of the control Hamiltonian $\partial_RH$ over a
constrained microcanonic ensemble; see (\ref{katu}, \ref{ab3}).  As
follows from (\ref{ab1}, \ref{ab3}), the speed of entropy reduction
decreases under information limitations: 
\BEA
|\frac{\sd S}{\sd \tau}|\leq
|\widetilde{\frac{\sd S}{\sd \tau}}|. 
\EEA
In particular, $\psi=\frac{\sd S}{\sd \tau}=0$ for $\Phi={\rm const}$.

{\it Examples}. We illustrate the obtained feedback schemes via the
celebrated example of adiabatic physics that is a harmonic oscillator
with the controlling frequency:
\BEA
H=\frac{p^2}{2}+\frac{Rq^2}{2}. 
\label{kobo}
\EEA
This Hamiltonian with the feedback controlling frequency is close to
the experimental situation realized in Ref.~\cite{cohen}.
In the context of adiabatic feedback control, the harmonic oscillator  
(\ref{kobo}) displays two interesting effects that exist as well in more elaborated
situations \cite{aa}: control without systematic motion of the
controlling parameter and qualitative changes in the control setup upon
information limitations. 
For a constant $R$ the period of the oscillator is $2\pi/\sqrt{R}$, and
the adiabatic approach applies at least for $1\ll\tau_R\sqrt{R}/2\pi$.
This is an ergodic system and Eq.~(\ref{ooo1}) implies for the
optimal entropy-reducing control
\BEA
\label{kaban}
F(E,R,q)=-\frac{\fm}{2}\,{\rm sign}(q^2-\frac{E}{R}). 
\EEA
Eq.~(\ref{kaban}) leads via
(\ref{hek1}, \ref{hek2}, \ref{quito}) to $R$ 
constant in the slow time (though $R$ is not constant on the fast time), 
\BEA
\label{canes}
\frac{\sd R}{\sd
\tau}=0, 
\EEA
exponential decay of energy (cooling),
\BEA
E(\tau)=E(0)e^{-\frac{\fm }{\pi R}\frac{\tau}{\tau_R}}, 
\EEA
and thus to linear decay of the entropy $S=\ln[\frac{2\pi
E}{\sqrt{R}}]$: $S(\tau)=S(0)-\frac{\fm}{\pi R}\,\frac{\tau}{\tau_R}$.
As intuitively expected, entropy reduction relates to cooling.  
Eq.~(\ref{canes}) shows that the controlling parameter $R$ does not move 
in average, i.e., on the slow time. The cooling is achieved due to the motion
of $R$ on the fast time-scale; see (\ref{2}, \ref{kaban}).

\begin{figure}
\includegraphics[width=6cm]{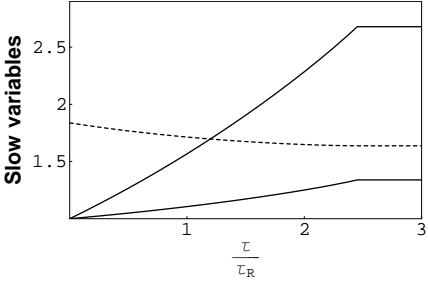}
\caption{
Entropy reduction scheme (\ref{campinas1}, \ref{campinas2}) for harmonic
oscillator. Slow variables versus dimensionless time $\tau/\tau_R$:
Energy $E$ (lower curve), frequency $R$ (upper curve), and entropy $S$
(dashed curve). For the parameters we take $\fm =1$ and $L=1$. The
curves saturate for $\tau/\tau_R>2.5$ when the available information is
not anymore sufficient for any entropy reduction. 
}
\hfill
\label{fig_1}
\end{figure}

To limit the information available about the coordinate $q$, we assume
that it is only known whether $q$ is larger or smaller than a positive
constant $L$. The function $\Phi(q)$ in (\ref{kentavr}) thus takes only two
distinct values, e.g., $\Phi(q)=\theta(L-q)$. This brings from
(\ref{karro}, \ref{kentavr}) a control setup
\BEA
\label{kaplan}
F(E,R,q)=\fm\,{\rm sign}(L-q)\, \theta(2E-RL^2).  
\EEA
The latter step function is there, since for small $E$ (or large $R$)
the oscillator is located next to $q=0$ and its position cannot be
utilized by the feedback. Thus for those values of $E$ it is best to do
nothing, $F=0$, since any control (under assumed information limits) will
increase the entropy (disorder). We get from (\ref{hek1}, \ref{hek2},
\ref{kaplan}) and from (\ref{kentavr}--\ref{ab3})
\BEA
\label{campinas1}
\frac{\sd R}{\sd \tau}=\frac{2\fm}{\tau_R\pi}\, \theta(1-\xi)\,
\arcsin\xi, \qquad  \xi\equiv\sqrt{\frac{RL^2}{2E}},
\\
\frac{\sd S}{\sd \tau}
\equiv \frac{\sd }{\sd \tau}
\ln[\frac{2\pi E}{\sqrt{R}}]
=-\frac{\fm}{\pi R\tau_R}
\, \theta(1-\xi)\,\xi\sqrt{1-\xi^2},
\label{campinas2}
\label{orojkhon}
\EEA
while the equation for $\frac{\sd E}{\sd \tau}$ can be recovered from
(\ref{campinas1}, \ref{campinas2}). The behavior of $E(\tau)$, $R(\tau)$
and $S(\tau)$ is shown in Fig. 1. We see that the entropy
reduction rate is not simply smaller than the optimal one, but it is
realized via energy increase (heating) rather than cooling. 

For the considered oscillator model, let us illustrate that
the Boltzmann expression is not the proper defintion of entropy for a
finite system. Recalling (\ref{entropy}, \ref{boltz}) and using
(\ref{orojkhon}) we get that for the harmonic oscillator (\ref{kobo}):
$S_B=\ln[\frac{2\pi}{\sqrt{R}}]$. It is seen that {\it i)} $S_B$ does
not depend on the energy, so that attempting to define the temperature via the
standard formula (\ref{dadd}) will lead to zero temperature, not a
reasonable result.  {\it ii)} $S_B$ is not adiabatic invariant, one can
decrease it via an open-loop control by changing $R$. 

{\it Adiabatic invariant.} Eq.~(\ref{kaplan}) provides a control setup,
where the dependence on the slow and fast variables factorizes:
\BEA
F(E,R,z)=g(E,R)\phi(z). 
\EEA
For this case the slow variable system
(\ref{hek1}, \ref{hek2}) possesses an integral of motion, i.e., an
adiabatic invariant. One deduces from (\ref{hek1}, \ref{hek2}):
\BEA
\frac{\sd }{\sd \tau}\int \d z\, \phi(z)\,\theta(E-H(z,R))=0. 
\EEA
This conservation generalizes to many-dimensional systems the adiabatic
invariance found in \cite{lisak}.  For $\phi(z)={\rm const}$, where
there is no feedback over fast variables, we are naturally 
back to the usual conservation of the phase-space volume. 

{\it In conclusion}, we developed an adiabatic approach for the adaptive
feedback control of Hamiltonian systems. It is derived assuming
ergodicity of two observables and thus applies to the most of chaotic
systems and some integrable ones.  The approach reduces the control
problem to two equations (\ref{hek1}, \ref{hek2}) describing the
evolution of slow variables.  With help of these equations we got a
general upper bound (\ref{ab2}) on the rate of negentropy (order)
production induced by the feedback control.  This bound is achieved for
discontinuous control field (\ref{ooo1}), and is related to the
fluctuation of the control Hamiltonian over the microcanonic ensemble.

The method describes the information-entropy trade-off: how the maximal
negentropy production rate decreases when the information available for
the feedback gets limited. The example of harmonic oscillator with the
controlling frequency shows that information limitations do change the
quanlitative features of the control dynamics. In particular, the
entropy reduction is realized via heating the system. Note that in the 
present approach we standardly modeled the information limitation via the
reduction of the data available to the controller.  

The Hamiltonian dynamics finds applications well beyond the proper
(statistical) mechanics, e.g., in hydrodynamics \cite{berdi} or in
ecology \cite{kerner}.  Control issues in these fields are well known,
and since our methods are not system-specific, they may apply to
controlling a vortex flow, or to optimizing the harvest production
\cite{aa}. 

\acknowledgements

We thank K.G. Petrosyan for useful discussions. 

The work was supported by Volkswagenstiftung (grant ``Quantum thermodynamics:
energy and information flow at nanoscale")
and by CRDF Grant No. ARP2-2647-YE-05.

\end{document}